\documentclass[11pt,a4paper]{article}
\pdfoutput=1 
\usepackage[utf8]{inputenc}
\usepackage{jheppub} 
\usepackage{graphicx}
\usepackage{bm}
\usepackage[T1]{fontenc} 

\usepackage{caption}
\usepackage{framed}
\usepackage{xcolor}

\usepackage[utf8]{inputenc}
\usepackage[cal=cm,scr=dutchcal]{mathalfa}
\title{\boldmath Strolling along gravitational vacua}

\author[a]{Emine Şeyma Kutluk,}
\author[b]{Ali Seraj,}
\author[a,c]{Dieter Van den Bleeken}
\affiliation[a]{Physics Department, Bo\u{g}azi\c{c}i University\\
 34342 Bebek / Istanbul, Turkey}
\affiliation[b]{Universit\'{e} Libre de Bruxelles and International Solvay Institutes\\ CP 231, B-1050 Brussels, Belgium}
\affiliation[c]{Secondary address\\
Institute for Theoretical Physics, KU Leuven\\
3001 Leuven, Belgium}

\emailAdd{emine.kutluk@boun.edu.tr}
\emailAdd{aseraj@ulb.ac.be}
\emailAdd{dieter.van@boun.edu.tr}

\abstract{We consider General Relativity (GR) on a space-time whose spatial slices are compact manifolds $M$ with non-empty boundary $\partial M$. We argue that this theory has a non-trivial space of `vacua', consisting of spatial metrics obtained by an action on a reference flat metric by diffeomorpisms that are non-trivial at the boundary. In an adiabatic limit the Einstein equations reduce to geodesic motion on this space of vacua with respect to a particular pseudo-Riemannian metric that we identify. We show how the momentum constraint implies that this metric is fully determined by data on the boundary $\partial M$ only, while the Hamiltonian constraint forces the geodesics to be null. We comment on how the conserved momenta of the geodesic motion correspond to an infinite set of conserved boundary charges of GR in this setup.}

\def\calg{{\mathscr G}}
\def\calk{{\cal K}}

\def\calsg{{\boldsymbol{\mathscr g}}}

\def\calh{{\cal H}}
\def\calm{{\mathcal{M}}}

\def\calo{{\cal O}}
\def\cals{{\cal S}}

\def\calv{{\cal V}}

\def\a{\alpha}
\def\b{\beta}
\def\g{\gamma}

\def\uk{{\underline{k}}}
\def\ul{{\underline{l}}}

\def\m{{\mu}}
\def\n{{\nu}}
\def\pd{\partial}
\def\de{\delta}
\def\rg{{\mathrm{g}}}

\def\Ad{{\mathrm{Ad}}}

\def\rn{{\mathrm{n}}}
\def\rt{{\mathrm{t}}}
\def\pM{{\partial M}}

\newcommand{\cL}{{\mathcal L}}
\newcommand{\cd}{{\nabla}}

\usepackage[normalem]{ulem}
\begin{document}
    \maketitle
    \flushbottom
    
    \section{Introduction}
    There is some merit in viewing (classical) field theory as a version of particle mechanics, with the particle moving on an infinite dimensional rather than finite dimensional manifold. This manifold -- the configuration space of time-independent fields -- comes equipped with a metric provided by the kinetic term of the field theory Lagrangian. In the case of Yang-Mills (YM) theory this point of view was concisely introduced in \cite{Babelon:1980uj} while for General Relativity (GR) this is the program of Geometrodynamics going back to \cite{Wheeler:1957mu}, see \cite{Orland:1996hm, Giulini:2009np} for an entry into the more recent literature. Making this simple idea precise and/or mathematically rigorous becomes rather involved, mainly due to the presence of gauge symmetries and the corresponding constraints. To simplify matters the spatial slices $M$ of space-time are typically assumed compact and without boundary, so that all gauge symmetries are equivalences that are quotiented out. In case $M$ has a non-empty boundary $\partial M$ there are global gauge symmetries -- gauge symmetries which are non-trivial at the boundary -- that are not to be interpreted as equivalences but rather as global symmetries connecting various physically inequivalent configurations. This is a feature that makes the setup richer but at first sight also more complicated. We would argue however that instead it opens up an unexplored corner -- that we refer to as the space of vacua -- which can be isolated and where things simplify dramatically, providing an extra handle on the geometric approach to field theory. We will now explain this more concretely and summarize our main results, while referring to the discussion in section \ref{discs} for further motivations and possible applications.
    We define a `vacuum' in our classical field theory setting as a (non-singular) equilibrium configuration of absolute minimal energy (in some particular static gauge).  The latter condition is needed to differentiate from possible (topological) solitonic solutions. Global symmetries of the theory can be broken and will then automatically generate a multitude of vacua. In the simplest cases all vacua are connected in this way so that the space of vacua becomes a homogeneous space $\calv=\cals/
    \calk$, with $\cals$ the group of global symmetries and $\calk$ the isotropy group formed by transformations that leave a vacuum invariant. This is the case in the simple example of a complex scalar field with Mexican hat potential, but also in pure YM theory \cite{Lechtenfeld:2015uka, Seraj:2017rzw} and -- as is the subject of this paper -- GR, when the spatial slices of space-time have a boundary. In these cases the space of vacua is infinite dimensional because the group of global symmetries is that of global gauge transformations -- itself isomorphic to the group of boundary gauge transformations -- while the isotropy group is finite dimensional. The space of vacua is a small subset of the configuration space and carries an induced geometry through this embedding.  A point we stress -- as it is one of the motivations for our work -- is that the space of vacua equipped with this metric has an important physical interpretation. In a limit of slow time dependence -- which we refer to as the adiabatic limit -- the dynamics of the full field theory simplifies and reduces to geodesic motion on the space of vacua. A similar limit can be considered around solitonic equilibria where it goes under the name of the Manton or moduli space approximation \cite{Manton:1981mp}. This present work analyzes this adiabatic dynamics and the associated geometry on the space of vacua for GR and is a continuation of our work \cite{Seraj:2017rzw} on YM. There are many similarities between the adiabatic dynamics in both theories, but also some interesting differences. Let us list our main results while outlining our paper.
    
    In section \ref{toy} we illustrate in two simple particle mechanics examples how the dynamics reduces to free motion on the space of vacua in an adiabatic approximation. Readers familiar with this topic (or the Manton approximation) can skip this section, although they might find the second example that shares some of the peculiarities of GR new and interesting. 
    
    In section \ref{setupsec} we set the stage for our analysis. We partially gauge-fix GR by going to Gaussian Normal Coordinates (GNC) -- also known as synchronous gauge -- which is the analog to temporal or static gauge in YM. We review how in this gauge the Einstein equations split into two constraints and a dynamic equation that can be obtained from a Lagrangian in natural form. We point out -- that due to the presence of a boundary-- the residual gauge invariances are boundary preserving spatial diffeomorphisms and then formally define the space of vacua in \eqref{vacdef}.
    
    Section \ref{mainsec} contains our analysis of the adiabatic approximation and original results. We show how the momentum constraint -- as does the Gauss constraint in YM -- eliminates (most) local diffeomorphisms and allows to rewrite the space of vacua and its metric purely in terms of data on the boundary $\partial M$. We present the metric on the space of vacua in \eqref{bndmetric2}. To find its actual value when evaluated on a boundary diffeomorphism one needs to solve a boundary value problem that we show is well defined and has a unique solution. The first interesting difference with YM is that this metric for GR has a mixed signature and is thus pseudo-Riemannian rather than Riemannian. This is all the better since we show that the Hamiltonian constraint -- in addition to removing the remaining local diffeomorphisms -- implies a constraint on the adiabatic motion that is equivalent to requiring the geodesics on the space of vacua to be null. We end the section by elaborating on the homogeneous space structure which allows to express the metric in terms of a fixed inner product and show how the conserved momenta of the geodesic motion correspond to an infinite set of charges associated to the boundary symmetries that can be independently derived using covariant phase space methods in full GR.
    
    In section \ref{spexample} we work out a specific example where the boundary  $\partial M$ is a round sphere and compute the metric rather explicitly, see \eqref{spmetric}.
    
    Finally, in section \ref{discs} we comment on the relation between our work and recent developments in the literature, mentioning opportunities for further research.
    
    There are three appendices -- \ref{techtime}, \ref{fol} and  \ref{apphodge} -- containing details on some technical material we use. 
     
\section{Illustration of the adiabatic approximation}\label{toy}
    Given a theory with time-independent -- i.e. equilibrium -- solutions it is natural to investigate if one can obtain new solutions by introducing a slow time dependence. One area where this has proven highly successful and interesting is that of topological solitons where the approximation method goes under the name of the Manton or moduli-space approximation \cite{Manton:1981mp}. There one starts with an infinite  dimensional field space and reduces the problem to that of motion on a finite dimensional subspace, that is often referred to as the moduli space. In our setup we start with an infinite dimensional field space and reduce the problem to motion on a smaller infinite dimensional moduli space which has the natural interpretation as the space of vacua. For a general discussion of the adiabatic/Manton approximation in field theory we refer to \cite{Manton:1981mp, Manton:2004tk, Weinberg:2006rq, Stuart:2007zz, Seraj:2017rzw}. The main idea of the method can however already be illustrated at the level of particle mechanics where one only deals with finite dimensional spaces, which is what we will do in this section.  The first example is absolutely elementary but sets the stage for a second example where we introduce time-reparametrization invariance to illustrate an important novelty that will appear also in our application of the method to GR: we will obtain a moduli space/space of vacua naturally equipped with a pseudo-Riemannian metric. Moreover, the Hamiltonian constraint -- a consequence of the time reparametrization invariance -- restricts the dynamics to motion along null curves.
        
    \subsection{A standard example}
    Consider a non-relativistic particle in 3 dimensions with a `Mexican hat' potential:
    \begin{equation}
    L=\frac{1}{2}\rg(\dot X,\dot X)-V(X)=\frac{m}{2}\delta_{IJ}\dot X^I\dot X^J-\frac{1}{2}(\delta_{IJ}X^IX^J-R^2)^2 \,.
    \end{equation}
    The solutions of absolute minimal energy are
    \begin{equation}
    \partial_t \bar X^I=0\quad \mbox{with} \quad \delta_{IJ}\bar{X}^I\bar X^J=R^2 \,. 
    \end{equation}
    We will refer to those $\bar X$ minimizing the potential as `vacua' and the equipotential surface generated by them as the space of vacua $\calv$, so that in this example $\calv=S^2$. What is special in this example -- and is also the case in GR and YM -- is that all vacua are connected by a transitive action of symmetries of the Lagrangian, which guarantees that the space of vacua is a homogeneous space, in this case $\calv=\mathrm{SO}(3)/\mathrm{SO}(2)$. Now consider a particle that is initially on the minimal potential surface but has a non-zero initial velocity. The result will be some non-trivial motion which is a combination of angular rotations, so called zero-modes, and motion `up the potential' so called normal modes. The adiabatic approximation then simply amounts to the insight that for small initial velocity the motion in the normal directions is of second order -- and therefore subleading -- and additionally decouples from the motion along the zero-mode directions, which becomes free -- i.e. geodesic. Let us spell this out in this simple example. We can change coordinates to
    \begin{equation}
    X^1=(R+\rho)\sin z^1\cos z^2 \, ,\qquad  X^2=(R+\rho)\sin z^1\sin z^2 \, ,\qquad  X^3=(R+\rho)\cos z^1 \, .
    \end{equation}
    The $z^\alpha$ parametrize the vacua $\bar X(z)$ and are hence coordinates on $\calv$, the coordinate $\rho$ describes the normal direction. The Euler-Lagrange equations now read
    \begin{eqnarray}
\label{eom1}    (R+\rho)\left(\ddot z^\alpha+\Gamma_{\b\g}^\alpha \dot z^\b\dot z^\g\right)+2\dot{\rho}\dot z^\alpha&=&0\,,\\
\label{eom2}    \ddot \rho+\frac{4}{m}\rho(\rho+2R)(\rho+R)-(R+\rho) g_{\a\b}\dot z^\a\dot z^\b=0 \, ,
    \end{eqnarray}
    where $\Gamma_{\alpha\beta}^\gamma$ is the Levi-Civita connection of $g_{\alpha\beta}$, the metric induced on $\calv=S^2$. Now note that these are coupled equations such that turning on $\dot z$ will source motion in $\rho$ through the second equation which then backreacts on the motion in $z$ through the $2\dot{\rho}\dot {z}$ term in the first equation. Let us now assume that velocities are small, i.e. all time derivatives are of order $\epsilon$. Allowing $z^\alpha$ to be arbitrarily large, i.e. of zeroth order, the second equation then however implies that $\rho$ needs to be small, i.e. of order $\epsilon^2$. This then makes the coupling term in the first equation of 4th order so that it can be ignored with respect to the first two. It implies that the motion along $\calv$ decouples from the normal motion and becomes purely geodesic. This result is generic to the adiabatic approximation and extends to field theory as well.

    \subsection{A non-standard example}
    The previous example captures the main features of the adiabatic approximation, but it doesn't have time reparametrization invariance. This is a key ingredient in GR and so it might be useful to illustrate this feature and its consequences in an equally elementary example as well.\\
\hfill \\    
Consider the Lagrangian\footnote{One can eliminate $N$ to recover a relativistic particle with a spacetime-dependent mass, as could appear through coupling to a scalar field.}
	\begin{equation}
	L=\frac{\eta_{\mu\nu}\dot X^\mu\dot X^\n}{2N}-N\frac{W(X)^2}{2} \,.\label{repth}
	\end{equation}
	Here $\eta_{\m\n}$ is the flat 4d Minkowski metric and the function $W$ could take various forms. We will make the simple and instructive choice
	\begin{equation}
	W=\eta_{\m\n}X^{\nu}X^{\mu}-R^2 \,.
	\end{equation}
	This theory \eqref{repth} has time reparametrizations as a gauge symmetry:
	\begin{equation}
	\delta X^\m=\dot X^\mu \epsilon \,, \qquad\delta \dot X^\mu=\ddot X^\mu\epsilon+\dot{\epsilon}\dot{X}^\mu \, , \qquad \delta N=\dot N\epsilon+\dot \epsilon N
	\end{equation}
	and its equations of motion are
	\begin{eqnarray}
	-\eta_{\m\n}\dot X^\mu\dot X^\nu&=&N^2 W^2\label{repeq1}\,,\\
	\eta_{\m \n} \dot{X}^{\n} \frac{\dot{N}}{N}-\eta_{\m\n}\ddot X^{\nu}&=&N^2W\partial_\mu W \, .\label{repeq2}
	\end{eqnarray}
	We can gauge-fix the reparametrization invariance by choosing $N=1$. In this gauge the equation \eqref{repeq2} becomes the Euler-Lagrange equation of a Lagrangian in natural form:
	\begin{equation}
	L=\frac{1}{2}\rg(\dot X,\dot X)-V(X)=\frac{1}{2}\eta_{\m\n}\dot X^\mu\dot X^\n-\frac{(\eta_{\m\n}X^{\nu}X^{\mu}-R^2)^2}{2}\,,\label{dSlag}
	\end{equation}
	while equation \eqref{repeq1} turns into the constraint
	\begin{equation}
	-\eta_{\m\n}\dot X^\mu\dot X^\nu=(\eta_{\m\n}X^{\nu}X^{\mu}-R^2)^2.\label{repcon}
	\end{equation}
	Note that the metric defining the kinetic term -- i.e. $\eta_{\m\n}$ -- is not positive definite. This is similar to the situation in GR and crucial for consistency as we will see.
	Again in this example the space of vacua, or minima of the potential, is generated by global symmetries of the theory making it a homogeneous space: 
	\begin{equation}
	\calv=\mathrm{SO}(1,3)/\mathrm{SO}(1,2)\,.
	\end{equation}
	A key and crucial difference with the first example is that in this case the space of vacua is three dimensional de Sitter space\footnote{See e.g. \cite{deWit:2002vz} for an introduction to homogeneous spaces and the exposition of some classic examples.} --  $\calv=dS_3$ --  which is a Lorentzian rather than Riemannian manifold. That indeed the metric on $\calv$ induced by the kinetic term is the natural one can be made explicit via a change of coordinates:
	\begin{eqnarray}
	X^0&=&(R+\rho)\sinh z^0 \, , \qquad X^1=(R+\rho)\cosh z^0\cos z^2\sin z^1 \, ,\\
	X^2&=&(R+\rho)\cosh z^0\sin z^2\sin z^1 \, , \qquad X^3=(R+\rho)\cosh z^0\cos z^1 \, .
	\end{eqnarray}
In these coordinates the Lagrangian \eqref{dSlag} takes the form
	\begin{equation}
	L=\frac{m}{2} \left( \rho + R \right)^2 g_{\a\b}\dot z^\a\dot z^\b+\frac{m}{2}\dot \rho^2- \left( \left( \rho + R \right)^2-R^2 \right)^2
	\end{equation}
	with
	\begin{equation}\label{dS metric}
	g_{\a\b}d z^\a d z^\b=-(dz^0)^2+\cosh^2z^0\left((dz^1)^2+\sin^2z^1 (dz^2)^2\right)
	\end{equation}
	The equations of motion of this Lagrangian read the same as \eqref{eom1} and \eqref{eom2} with the metric and Christoffel symbol replaced by those of de Sitter space \eqref{dS metric}. 	We use notation similar to the previous example and the main text to indicate that the $z^\alpha$ parametrize $\calv$ while $\rho$ parametrizes the normal direction. The adiabatic approximation again amounts to taking $\dot z=\calo(\epsilon)$, which implies through the equations that $\rho=\calo(\epsilon^2)$. Accordingly the equations of motion reduce to geodesic motion on $\calv$ with respect to the induced metric $g_{\alpha\beta}$, while the constraint \eqref{repcon} becomes
	\begin{equation}
	g_{\a\b}\dot z^\a\dot z^\b=0 \,.
	\end{equation}
	Note that the constraint originating from time reparametrization invariance enforces the geodesic motion on $\calv$ to be null. That this is possible in the first place is due to the pseudo-Riemannian nature of $\calv$ in such theories. We will see that GR shares both these features.

    \section{Setup, notations and conventions for GR}\label{setupsec}
    In this section we set the stage for our main analysis in the next section.

    \subsection{General relativity in Gaussian normal gauge}
   The theory we will consider is standard Einstein gravity, i.e. general relativity (GR). For simplicity we will restrict our discussion to vanishing cosmological constant and 4 dimensions, with space-time of the form $\mathbb{R}\times M$, but we expect our results to generalize to these other cases as well. A key ingredient in our discussion is that we take the spatial manifold $M$ to be a compact manifold with non-empty boundary $\partial M$. Again for simplicity we assume the topology of $M$ to be trivial, i.e. we consider $M$ to be homeomorphic to the closed 3-ball. The dynamics are described by the Einstein-Hilbert action
   \begin{equation}
   S_{\mathrm{GR}}=\int_{\mathbb{R}\times M} d^4x\,\sqrt{-g}\, R^{(4)} \, .
   \end{equation}
   Here $R^{(4)}$ is the Ricci scalar\footnote{We explicitly indicate that this is the 4d Ricci scalar of the 4d metric $g$, since in most of the paper $R$, the 3d Ricci scalar of the spatial metric $h$ will appear.} associated to the metric $g_{\mu\nu}$. The equations of motion are
   \begin{equation}
   R_{\mu\nu}^{(4)}=0 \, .\label{Eeq}
   \end{equation}
   We now split the coordinates $x^\mu=(t,x^i)$ and choose them such that the metric is in Gaussian normal form\footnote{Note that while any metric can locally be expressed in Gaussian normal coordinates (GNC), there can be obstructions to extending these coordinates globally, due to possible crossings of the time-like geodesics used in the definition of the coordinates. For the special class of space-times that we consider -- outlined at the beginning of this section -- this amounts to a restriction of the validity of these coordinates to a possibly short but finite time interval. The adiabatic motion we will consider takes place -- by definition -- on a time scale $\tilde t$ that is parametrically small as measured in units of the time $t$ of the GNC, i.e. $t=\epsilon \tilde t$. So we expect there to be plenty of room for non-trivial adiabatic motion inside the time interval on which the GNC are well-defined. It would be interesting to compare in an actual solution the time scale on which the adiabatic approximation exceeds a certain error to the scale on which the GNC break down inside a certain spatial volume, but this is left for future work.}:
   \begin{equation}
   ds^2=g_{\mu\nu}dx^\mu dx^\nu=-dt^2+h_{ij}(t,x)dx^idx^j \, .\label{mGNC}
   \end{equation}
   We can think of this choice of coordinates as a (partial) gauge fixing of the  diffeomorphism invariance of the theory. The Gaussian normal coordinates (GNC) are not unique. The form of the metric \eqref{mGNC} remains invariant under two types of coordinate transformations. First there are the spatial diffeomorphisms 
   \begin{equation}
   \xi^0(t,x)=0\qquad \xi^i(t,x)=\chi^i(x)\label{spatdif}
   \end{equation}
   and additionally there are `local boosts'\footnote{In the special case $h_{ij}=\delta_{ij}$, $\alpha(x)=v_ix^i$ and they thus reduce to a standard Lorentz boost.}:
   \begin{equation}
  \xi^0(t,x)=\alpha(x)\qquad \xi^i(t,x)=\partial_j\alpha(x)\int^{t}_{t_0} h^{ij}(t',x)dt' \, .\label{alpha}
   \end{equation}
   These boosts play an important and subtle role in the canonical formulation of gravity-- see e.g. \cite{Isham:1984sb, Isham:1984rz} -- but will not enter our analysis as we explain below.
   
   In the choice of gauge \eqref{mGNC} the Einstein equations \eqref{Eeq} take the standard form of the initial value formulation
   \begin{eqnarray}
   \cd^i \left( \dot h_{ij}- h^{kl}\dot h_{kl} h_{ij} \right) &=&0 \, , \label{const1} \\
   R+\frac{1}{2}h^{ij}h^{kl}\dot h_{i[j}\dot h_{k]l}&=&0\label{const2} \, ,\\
   \ddot{h}_{ij} + h^{kl}(\dot h_{ik} \dot h_{jl}-\dfrac{1}{2}\dot h_{ij} \dot h_{kl})-2{R}_{ij}&=&0 \, . \label{dynamicaleq}
   \end{eqnarray}
   The first two of these equations are constraints, referred to as the momentum constraint \eqref{const1} and Hamiltonian constraint \eqref{const2} respectively. As we will see, both will have important but conceptually different roles to play in our construction. The third equation \eqref{dynamicaleq} contains the second order time derivatives and is thus the dynamical equation. It  is the Euler-Lagrange equation\footnote{To have a well defined variational principle in the presence of a boundary one has to impose boundary conditions on the fields and/or variations which in this case are $\left.\left(n_i\cd_j(h^{i[j}h^{k]l}\de h_{kl})\right)\right|_{\partial M}=0$, with $n^i$ the unit normal to the boundary. Under these conditions our discussion is self-consistent. It might be interesting to investigate a possible boundary contribution to \eqref{natlag} that could change the variational principle. Let us stress that in our setup it is crucial that the boundary metric is left free to fluctuate.} of an action in natural form:
   \begin{equation}
   S_\mathrm{nat}=\int dt\, \frac{1}{2}\rg(\dot h,\dot h)-V(h) \, .\label{natlag}
   \end{equation}
   The metric appearing in the kinetic term is the Wheeler-deWitt (WdW) metric 
   \begin{equation}
   \rg(\delta_1h,\delta_2h)=\frac{1}{2}\int_M d^3x \sqrt{\det h} h^{i[k}h^{j]l}\delta_1 h_{ij}\delta_2h_{kl} \, ,\label{WdW}
   \end{equation}
   while the potential is determined by the Ricci scalar
   \begin{equation}
   V(h)=-\frac{1}{2}\int_M d^3x\,\sqrt{\det h}\,R(h) \, .\label{pot}
   \end{equation} 
   
   This way to formulate the theory identifies a (pre-)configuration space $\calm$ of time-independent spatial metrics $h_{ij}(x)$, and a metric \eqref{WdW} on it. The dynamical 4d metrics on space-time in turn are identified -- through \eqref{mGNC} -- with motions, i.e. curves $h_{ij}(t,x)$ in this configuration space. The above observations, and various subtleties related to it, can be made more mathematically formal using the notion of superspace, see e.g. \cite{Giulini:1993ct}.
   
   The coordinate transformations \eqref{spatdif} and \eqref{alpha} that leave the Gaussian normal gauge \eqref{mGNC} invariant are also invariances of the Lagrangian \eqref{natlag}. Note however a conceptual difference between these two types of transformations in the aforementioned formulation: while the spatial diffeomorphisms \eqref{spatdif} have an action on the configuration space of the theory the local boosts \eqref{alpha} only act on the set of motions, or curves, in the configuration space. This distinction is analogous to that between translations and Galilean boosts in the theory of a non-relativistic particle. Our goal in this paper is to study slow motion generated by symmetries acting on the zero potential subspace of configuration space. It follows that we can restrict attention to the spatial diffeomorphisms and ignore\footnote{Note that the local boosts can be used to construct exact fast motion solutions, something of possible interest but outside the scope of this paper.} the local boosts. 
   Being a bit more precise, the variation of the Lagrangian \eqref{natlag} under a spatial diffeomorphism $\chi$ is
   \begin{equation}
   \delta L=\frac{1}{2}\oint_{\partial M}d^2y\, n_i \chi^i \sqrt{h}(h^{i[k}h^{j]l}\dot h_{ij}\dot h_{kl}+R) \, ,
   \end{equation}
   where $n$ is the unit normal to $\partial M$. It follows that only boundary preserving diffeomorphisms are invariances of the theory, these are the spatial diffeomorphisms whose generating vector field has vanishing normal component at the boundary:
   \begin{equation}
   \left.\chi^\perp\right|_{\partial M}=0\qquad\qquad (\chi^\perp=n_i\chi^i) \, .
   \end{equation}
   In the remainder of this paper \textit{we will often simply refer to boundary preserving spatial diffeomorphisms as `diffeomorphisms'}.  Let us denote the action of such a (inverse) diffeomorphism $\phi$ on the spatial metric $h$ with $\phi\cdot h$, which explicitly reads in terms of coordinates
   \begin{equation}
   (\phi\cdot h)_{ij}(x) = \frac{\partial \phi^k}{\partial x^i}\frac{\partial \phi^l}{\partial x^j}h_{kl}(\phi(x)) \, .\label{difdef}
   \end{equation}
   Here the right hand side involves the expression of the diffeomorphism as a coordinate transformation $\phi^i(x)\mapsto x^i$. Infinitesimally, when $\phi^i(x)=x^i+\chi^i(x)$ one has the standard expression
   \begin{equation}
   \delta_\chi h_{ij}=(\phi\cdot h)_{ij}(x)-h_{ij}(x)=(\nabla_\mathrm{s}\chi)_{ij} \, ,\qquad (\nabla_\mathrm{s}\chi)_{ij}\equiv L_\chi h_{ij}=\cd_i\chi_j+\cd_j\chi_i \, .
   \end{equation}
   
   \subsection{Vacuum metrics and the space of vacua}
   In analogy with particle mechanics and our analysis of Yang-Mills (YM) theory \cite{Seraj:2017rzw} we define a vacuum (spatial) metric $\bar h(x)$ as one for which the potential energy \eqref{pot} is extremal, which is equivalent to its Ricci curvature vanishing:
   \begin{equation}
  \qquad R_{ij}(\bar h)=0 \, .\label{ricflat}
   \end{equation}
   Note that vanishing potential is -- contrary to the situation in YM theory -- not equivalent to absolute minimal potential energy, since the \eqref{pot} is not bounded from below. Still -- and this is the key property -- it follows from the equations of motion that this is the unique condition that allows for equilibrium -- i.e. time independent -- solutions:
   \begin{equation}
   h_{\mathrm{eq}}(t,x)=\bar h(x) \, .\label{timeindep}
   \end{equation}
   The conditions \eqref{ricflat} and \eqref{timeindep} together with \eqref{mGNC} define a special set of static 4d Lorentzian metrics that solve the vacuum Einstein equations, but we stress that our notion of `vacuum metric' as defined above is much stronger. Indeed, it is well known that -- because the spatial manifold is 3 dimensional -- any (non-singular) solution to \eqref{ricflat} is diffeomorphic to the flat Euclidean metric\footnote{We assume $M$ to be topologically trivial, i.e. homeomorphic to the closed 3-ball.}:
   \begin{equation}
   \bar h=\phi\cdot \bar h_o \quad \mbox{where} \quad \bar h_{o\,_{ij}}=\delta_{ij} \, .\label{vacdef}
   \end{equation}
   One might therefore be tempted to treat all vacuum metrics as physically equivalent. Due to the presence of a boundary however, one should not identify metrics related by global diffeomorphisms, i.e. diffeomorphisms that are non-trivial on the boundary. This piece of lore is independently confirmed by a careful analysis of the equations (\ref{const1}-\ref{dynamicaleq}), as we argue in detail in the next section. It then follows from \eqref{vacdef} that the space of vacua, the set of physically inequivalent vacuum metrics, is the homogeneous space
   \begin{equation}
   \calv=\cals/\calk=\calg_0\backslash \calg/\calk \, .\label{homspac1}
   \end{equation}
   Here $\calg=\mathrm{Diff}(M,\partial M)$ is the group\footnote{Strictly speaking we consider only the connected component containing the identity.} of boundary preserving spatial diffeomorphisms introduced above, $\calg_0$ the normal subgroup of spatial diffeomorphisms that reduce to the identity on the boundary and $\cals=\calg/\calg_0=\mathrm{Diff}(\partial M)$ the quotient group which is isomorphic to the group of boundary diffeomorphisms. The group $\calk$ is that of boundary preserving spatial isometries -- namely those diffeomorphisms that act trivially on the reference metric $\bar h_o$ -- that form the isotropy group of the transitive action \eqref{vacdef}.  The result \eqref{homspac1} is fully analogous to that in the case of YM theory \cite{Seraj:2017rzw}. In the next section we will explain this characterization of the space of vacua $\calv$ in more detail and most importantly we will also construct a new left-invariant metric on this space.
   
   \section{Adiabatic motion on the space of vacua}\label{mainsec}
   The adiabatic method, or Manton approximation \cite{Manton:1981mp, Stuart:2007zz}, relies on a decoupling of the motion along an equipotential surface from the motion orthogonal to such a surface, in the limit of slow velocity. The motion along the equipotential surface is described fully non-linearly, as geodesic motion with respect to an induced metric, while the normal motion can be neglected in this approximation. Since symmetries generate equipotential surfaces they naturally lead to such adiabatic solutions. This includes gauge-symmetries, as famously exemplified by the interpretation of electric charge for non-abelian monopoles as motion in a gauge direction, see \cite{Weinberg:2006rq} for a pedagogic review. In this section we will investigate this method in the case of GR, by introducing time dependence into the spatial diffeomorphisms \eqref{difdef} and acting with them on a vacuum metric \eqref{vacdef}. This will essentially be a repetition of our work on Yang-Mills theory \cite{Seraj:2017rzw}, but as we will see there are some crucial and interesting differences.
   
   We start by considering a multi-parameter set of boundary preserving diffeomorphisms $\phi_z$. Via \eqref{vacdef} the parameters $z^\alpha$ will provide a complete set of coordinates on the space of vacua:
   \begin{equation}\label{intro-timedep}
   \bar h(z)=\phi_z\cdot \bar h_0 \, .
   \end{equation}
   Motion along the set of vacua $\calv$ can then be described by restricting time dependence of the spatial metric $h$ to be only through the parameters $z^\alpha$:
   \begin{equation}
   h(t,x)=\bar h(x;z(t)) \, .\label{mantonansatz}
   \end{equation}
   Adiabatic motions are then such metrics \eqref{mantonansatz} that solve the equations (\ref{const1}-\ref{dynamicaleq}) in the limit of small velocities $\dot z^\alpha$.  Note that for all metrics of the type \eqref{mantonansatz} the Ricci curvature vanishes, so the only quantity appearing in the constraints and equations of motion are the time derivatives of $h$, which are computed to be
   \begin{equation}\label{tangent vectors}
   \boxed{ \dot h=\delta_{\chi_z}\bar h={\bar \nabla}_\mathrm{s}\chi_z\qquad \chi_z^i=\dot \phi_z^{\uk} \phi_{z\,\uk}^i \, .}
   \end{equation}
   In words the formula above simply states that the velocity $\dot{h}$ is tangent to the space of vacua and thus takes the form of an infinitesimal diffeomorphism. The explicit expression of the vector field $\chi_z$ in terms of the diffeomorphism $\phi_z$ is derived in appendix \ref{techtime} and will play a role in the analysis of section \ref{homspac}.
   Inserting \eqref{tangent vectors} into the WdW metric \eqref{natlag} will lead to the induced metric on the space of vacua and the dynamics is that of a particle moving on an infinite dimensional curved space:
   \begin{equation}
   S[z(t)]=\int dt \frac{1}{2}\bar \rg_z(\dot z,\dot z)\qquad \bar \rg_z(\dot z,\dot z)=\rg(\bar \nabla_\mathrm{s}\chi_z,\bar\nabla_\mathrm{s}\chi_z) \, .\label{action1}
   \end{equation}
   The Euler-Lagrange equations for this theory are geodesics with respect to $\bar \rg_z$, which provide adiabatic solutions to \eqref{dynamicaleq} through \eqref{mantonansatz}.
   In interpreting the expression above one should not forget that there are also constraints in the theory and the metric above should only be evaluated for velocities $\nabla_\mathrm{s} \chi_z$ satisfying (\ref{const1}, \ref{const2}). In the following subsections we will analyze and formally solve these two constraints in turn. As we will see they play conceptually rather different roles.
   
   \subsection{The momentum constraint and reformulation as a boundary theory}\label{momsec}
   For notational convenience we will suppress in this section  all bars and $z$ subscripts and consider it understood that all metrics appearing are vacuum metrics that depend on the parameters $z$. 
   We start our analysis by inserting \eqref{tangent vectors} in the momentum constraint \eqref{const1}. Using the fact that the Ricci tensor of \eqref{mantonansatz} vanishes, we can interchange covariant derivatives to rewrite \eqref{const1} as
   \begin{equation}
   \nabla^i\partial_{[i}\chi_{j]}=0 \, .\label{const1vec}
   \end{equation}
   We can now use this explicit form of the momentum constraint to further simplify \eqref{action1}. There is a remarkable interplay between the momentum constraint \eqref{const1vec} and the explicit form of the WdW metric such that upon integration by parts the bulk term vanishes. This allows to rewrite \eqref{action1}, when evaluated on solutions of \eqref{const1vec}, as a boundary integral\footnote{Note that although not manifest the RHS of \eqref{bndmetric1} is symmetric, as can be made manifest by performing the integration by parts in a symmetric way.}:
   \begin{eqnarray}
  \rg(\nabla_\mathrm{s}\chi^{(1)},\nabla_\mathrm{s}\chi^{(2)})=\oint_{\partial M} \sqrt{k}\,d^2y \left( \chi^a_{(1)} D^\perp \chi_a^{(2)}-K_{ab}\chi^a_{(1)}\chi^b_{(2)} \right)\, .\label{bndmetric1}
   \end{eqnarray}
   In the above expression we treated $\partial M$ as a hypersurface in a radial foliation of $M$, with the metric $h$ on $M$ inducing a metric $k$ on $\partial M$ that has extrinsic curvature $K_{ab}$. We denote the normal to the boundary as $n^i$ and choose coordinates $y^a$ on the boundary,  so that we can decompose 
   \begin{equation}
   \chi^i=\chi^\perp n^i+\chi^a e_a^i \, , \qquad\quad e^i_a\equiv\frac{\partial x^i}{\partial y^a} \, .
   \end{equation}
   We are working with boundary preserving diffeomorphisms so that 
   \begin{equation}
   \left.\chi^\perp\right|_{\partial M}=0 \, .\label{boundcond}
   \end{equation}
   Finally we introduced a normal covariant derivative which is compatible with the boundary metric:
   \begin{equation}
   D^\perp\chi^a\equiv n^i\partial_i\chi^a+\Gamma^a_b\chi^b \, .\label{normderv}
   \end{equation} See appendix \ref{fol} for further details on this and other foliation related definitions and formulae we use. 
   
   \subsubsection*{Bulk reconstruction from boundary data} Through \eqref{bndmetric1} we see that the metric on the space of vacua reduces to a boundary integral. However, it involve the normal derivative $D^\perp$ which involves how the diffeomorphisms extend into the bulk near the boundary. Here we show a remarkable result, namely that the momentum constraint restricts the bullk extension such that the metric is uniquely defined only in terms of boundary vector fields. In other words, we can think of \eqref{bndmetric1} as a metric on the space of boundary diffeomorphisms. To see this, consider a boundary vector field $\zeta^a(y)$. To be able to define the normal derivative $D^\perp$ of this vector field to compute \eqref{bndmetric1} one needs an extension $\chi^i(x)$ of the boundary vector fields into the bulk such that
   \begin{equation}
   \left.\chi^a\right|_{\partial M}=\zeta^a\label{bcond2}\,.
   \end{equation} One can then define $D^\perp\zeta^a\equiv\left.(D^\perp\chi^a)\right|_{\partial M}$ if one can specify a bulk extension for which this normal derivative is unique. We do this by demanding that the bulk extension satisfies the momentum constraint \eqref{const1vec} and the additional boundary condition \eqref{boundcond}. What complicates matters is that such a bulk extension is not unique, since the addition of an arbitrary exact part $\chi_i\to\chi_i+\partial_j\phi$ does not affect \eqref{const1vec} and can be chosen to respect \eqref{boundcond} as well. It is now crucial that this arbitrariness\footnote{Once the Hamiltonian constraint is taken into account as well the arbitrariness of the exact part will be fixed, as discussed in the next subsection.} of the bulk extension  does not affect the value of the normal derivative $D^\perp$ on the boundary. Showing this is technical and hence we will only outline the argument here and refer to appendix \ref{apphodge} for the mathematical background and a precise derivation.
   
   Given a boundary vector field/one-form $\zeta_a(y)$ one can compute its (boundary) exterior derivative $(d\zeta)_{ab}=2\partial_{[a}\zeta_{b]}$. As reviewed in appendix \ref{apphodge} there now exists a unique bulk co-exact/divergence-free one-form $\eta_i=\nabla^j\beta_{ji}$  such that
   \begin{equation}
   \left.\eta^\perp\right|_{\partial M}=0\,,\qquad \left.\partial_{[a}\eta_{b]}\right|_{\partial M}=\partial_{[a}\zeta_{b]}\,,\quad\mbox{and}\quad \nabla^i\partial_{[i}\eta_{j]}=0\,.\label{etaprop}
   \end{equation} One then notes that by the second equality $\zeta_a-\left.\eta_a\right|_{\partial M}$ is closed and hence -- by our assumption that the boundary is topologically a sphere -- (boundary) exact. It follows that we can uniquely decompose any boundary vector field/one-form as
   \begin{equation}
   \zeta_a=\left.\eta_a\right|_{\partial M}+\partial_a\sigma \, .\label{split}
   \end{equation}
   Using this decomposition we can now unambiguously define the `normal derivative' of a boundary vector field:
   \begin{equation}
   D^\perp\zeta_a\equiv\left.(D^\perp\eta_a)\right|_{\partial M}-K_a{}^{b}\partial_b\sigma \, .\label{bndDperp}
   \end{equation}
   Here on the right hand side $D^\perp$ is the normal derivative \eqref{normderv} evaluated on the boundary-tangent components of the bulk one-form $\eta_i$. Using the above definition we can now define an ($h$-dependent) inner-product on the space of boundary vector fields $\mathfrak{X}(\partial M)$:
   \begin{equation}
  \boxed{ \langle \zeta_{(1)}, \zeta_{(2)}\rangle_h\equiv\oint_{\partial M} \sqrt{k}\,d^2y \left( \zeta^a_{(1)} D^\perp \zeta_a^{(2)}-K_{ab}\zeta^a_{(1)}\zeta^b_{(2)} \right) \,.\label{bndmetric2}}
   \end{equation}
   
   We might need to clarify how \eqref{bndDperp} is indeed the normal derivative as computed via a suitable bulk extension $\chi_i$ of $\zeta_a$. This follows by considering the Helmholtz or Hodge-Morrey-Friedrichs decomposition\footnote{See appendix \ref{apphodge} and \cite{Schwarz:1995}.} of $\chi_i$:
  \begin{equation}
  \chi_i=\eta_i+\partial_i\phi \, .\label{chidecomp}
  \end{equation}
  Given this decomposition one can verify that the conditions \eqref{const1vec}, \eqref{boundcond} and \eqref{bcond2} are equivalent to \eqref{etaprop} and the following:
  \begin{equation}
  \left.(n^i\partial_i \phi)\right|_{\partial M}=0\,,\qquad \left.\partial_a \phi\right|_{\partial M}=\partial_a\sigma\,.\label{phicond}
  \end{equation}
  
  As we pointed out in the beginning of our argument, apart from the above two conditions the function $\phi$ is completely free. Interestingly all functions satisfying \eqref{phicond} share the same normal derivative of their exterior derivative:
  \begin{equation}
  \left.D^\perp \partial_a\phi\right|_{\partial M}=-K_{a}{}^b\partial_b\sigma \, .\label{phider}
  \end{equation}
  This follows because (see appendix \ref{fol} for details)
  \begin{equation}
  \left.\left(\partial_a(n^i\partial_i \phi)\right)\right|_{\partial M}=0\quad\Rightarrow\quad \left.(D^\perp\partial_a \phi)\right|_{\partial M}=-\left.K_a{}^b\partial_b\phi\right|_{\partial M}\,.\label{exrel}
  \end{equation}
  In summary \eqref{chidecomp} together with \eqref{phider} and the definition \eqref{bndDperp} imply that for any bulk extension $\chi_i$ of $\zeta_a$ satisfying \eqref{const1vec} and (\ref{boundcond}, \ref{bcond2}) one gets
  \begin{equation}
  D^\perp\zeta_a=\left.(D^\perp\chi_a)\right|_{\partial M}\,.
  \end{equation}
  The final upshot of this identification is that the metric \eqref{bndmetric1} for bulk diffeomorphisms satisfying \eqref{const1vec} and the metric \eqref{bndmetric2} for boundary diffeomorphisms are equal and independent of the part of $\chi_i$ not determined by the boundary conditions (\ref{boundcond}, \ref{bcond2}). The metric \eqref{bndmetric2} can be directly interpreted as an inner-product on the tangents space of the space of vacua $\calv$ at the point $h$, something which we will analyse further in section \ref{homspac}.
  
  \subsubsection*{Comments on signature}
  It is well known that the WdW metric \eqref{WdW} has a mixed signature, that actually jumps between different regions in superspace \cite{Giulini:1993ct}. As we will now show the metric \eqref{bndmetric2} on the space of vacua has a mixed -- but constant -- signature.
  
  The split \eqref{split} defines a direct product decomposition of the space of boundary vector fields:
  \begin{equation}
  \mathfrak{X}(\partial M)=\mathfrak{X}_+(\partial M)\oplus \mathfrak{X}_-(\partial M)\, .\label{dirprod}
  \end{equation}
  The subspace $\mathfrak{X}_+(\partial M)\equiv\{\zeta\in\mathfrak{X}(\partial M)\,|\, \zeta_a=\left.\eta_a[\zeta]\right|_{\partial M}\}$ is that of boundary vector fields that when extended into the bulk -- via the procedure explained above -- lead to a co-exact bulk vector, while $\mathfrak{X}_-(\partial M)\equiv\{\zeta \in\mathfrak{X}(\partial M)\,|\, \left.\eta_a[\zeta]\right|_{\partial M}=0\}$ are those boundary vector fields that get extended into exact bulk vector fields. Note that it follows from $\eqref{split}$ that elements of $\mathfrak{X}_-(\partial M)$ are also exact on the boundary, but that elements of $\mathfrak{X}_+(\partial M)$ are not necessarily co-exact on the boundary since $\eta$ can have an exact part\footnote{An exceptional case is when the boundary is the round 2-sphere where it happens that $\eta$ is always co-exact.}. 
  The key observation is now that -- under the assumption of convexity of the boundary --
  \begin{equation}
  \pm\langle\zeta,\zeta\rangle_h\geq 0\qquad\mbox{when}\quad\zeta\in \mathfrak{X}_\pm(\partial M)\,.\label{mixed}
  \end{equation}
  The first of these conditions follows because via \eqref{exrel} one computes that
  \begin{equation}
  \langle\zeta,\zeta\rangle_h=-2\oint_{\partial M} d^2y \sqrt{k} K^{ab}\zeta_a\zeta_b \qquad \mbox{when}\quad \zeta\in \mathfrak{X}_-(\partial M)
  \end{equation}
  and convexity of the boundary is equivalent to positive definiteness of its extrinsic curvature. The second condition is obtained by rewriting the metric in terms of the WdW metric via the bulk extension and using the co-exactness/divergencelessness to find
  \begin{equation}
  \langle\zeta,\zeta\rangle_h=\int_{M} d^3x \sqrt{h} h^{ik}h^{jl}\nabla_{(i}\chi_{j)}\nabla_{(k}\chi_{l)} \qquad \mbox{when}\quad \zeta\in \mathfrak{X}_+(\partial M) \, .
  \end{equation}
  We should stress that the decomposition \eqref{dirprod} is in general not orthogonal. This is most easily seen by computing that:
  \begin{equation}
  \langle\zeta^{(1)},\zeta^{(2)}\rangle_h=2\oint_{\partial M} d^2y \sqrt{k} \,\sigma^{(1)}D_a(K^{ab}\zeta_b^{(2)}) \qquad \mbox{when}\quad \zeta^{(1)}_a=\partial_a\sigma^{(1)}\in \mathfrak{X}_-(\partial M) \, .
  \end{equation}
  So we see that the orthogonal complement of $\mathfrak{X}_-(\partial M)$ is given by all boundary vectors that satisfy the condition $D_a(K^{ab}\zeta_b)=0$, which is not necessarily the case for elements of $\mathfrak{X}_+(\partial M)$. An important exception is again the round spherical boundary, see section \ref{spexample}.
  
  \subsubsection*{Summary}
  Let us here summarize the outcome of the somewhat technical discussion we made in this subsection. The main point is that implementation of the momentum constraint \eqref{const1} allows one to re-express the Lagrangian \eqref{action1} fully in terms of the boundary values $\zeta$ of the diffeomorphisms $\chi$:
  \begin{equation}
  S[z(t)]=\int dt \frac{1}{2}\bar \rg_z(\dot z,\dot z)\qquad \bar \rg_z(\dot z,\dot z)=\langle\zeta_z,\zeta_z\rangle_{\bar h(z)}\,.\label{action2}
  \end{equation}
  Furthermore the inner product $\langle\cdot,\cdot\rangle_h$, see \eqref{bndmetric2}, that appears is fully determined in terms of data on $\partial M$ only. This indicates that $z$ is a coordinate on (a quotient of) the space of boundary diffeomorphisms, since the velocities $\dot z$ can be identified with boundary vector fields. We will make this precise in section \ref{homspac}. Still two puzzles remain: 1) the metric \eqref{bndmetric2} has mixed signature and hence different types of geodesics, which ones describe the adiabatic solutions to GR? 2) given a boundary vector field, the bulk solution to the momentum constraint was only unique up to an exact part, what bulk vector is the one determining the adiabatic solution? 
  
  In particular the second question is important for conceptual reasons, since to construct the actual bulk metric that provides a(n approximate) solution to the Einstein equations one needs the full bulk diffeomorphism. As we discuss in the next subsection both issues are resolved by implementing the Hamiltonian constraint \eqref{const2}.

   \subsection{The Hamiltonian constraint and null geodesics}
   In the previous subsection we solved the momentum constraint \eqref{const1} on tangent vectors of the type \eqref{tangent vectors}. One key result was that in the Helmholtz decomposition \eqref{chidecomp} the co-exact/divergenceless part $\eta_i$ is fully determined in terms of the boundary data, while $\partial_i\phi$ remains undetermined. The next step is to additionally evaluate the Hamiltonian constraint \eqref{const2} on \eqref{tangent vectors}. Using that the vacua are Ricci flat it reads:
   \begin{equation}
   (\cd_i\chi^i)^2-\cd_{(i}\chi_{j)} \cd^i\chi^j=0 \, .\label{const2vec}
   \end{equation}
   Using \eqref{chidecomp} we can interpret the above equation as an equation for $\phi$:
   \begin{equation}
   (\cd_i\partial^i\phi)^2-\cd_{i}\partial_{j}\phi \cd^i\partial^j\phi-2\cd^i\eta^j\cd_{i}\partial_{j}\phi =\cd_{(i}\eta_{j)} \cd^i\eta^j\, .\label{phieq}
   \end{equation}
   The above equation is to be solved with $\eta_i$ given by the procedure of the last subsection, and under the boundary conditions
   \begin{equation}
   \left.n^i\partial_i\phi\right|_{\partial M}=0 \, ,\qquad \left.\partial_a\phi\right|_{\partial M}=\partial_a\sigma\,. 
   \end{equation}
   This non-linear differential problem appears to be non-trivial. The part quadratic in $\phi$ is the 2-Hessian operator \cite{Wang:2009} but the additional term linear in the second derivatives of $\phi$ makes \eqref{phieq} different from equations typically studied in the literature on Hessian equations. We feel confident however to conjecture that a non-singular solution for $\phi$ exists and is unique (provided the integrability condition \eqref{nullcond} discussed below is satisfied). Together with the results of the last subsection this conjecture is equivalent to the statement that for every boundary vector field $\zeta^a$ there is a unique extension into a bulk vector field $\chi$ with vanishing normal component at the boundary that solves both \eqref{const1vec} and \eqref{const2vec}.
   
   An important property of \eqref{const2vec} is that -- with the help of \eqref{const1vec} -- it can be rewritten as a conservation law:
   \begin{equation}
   \nabla_i j^i=0\label{conseq}
   \end{equation}
   where
   \begin{eqnarray}
   j^i&=&\chi_k \cd^{(i}\chi^{k)}-\chi^i\cd_k\chi^k\label{current}\\
   &=&-2\partial^{[i}\phi\cd^{k]}\partial_k\phi-2\eta^{[i}\cd^{k]}\partial_k\phi+\cd^{(i}\eta^{k)}\partial_k\phi+\eta_k \cd^{(i}\eta^{k)} \, .
   \end{eqnarray}
   Integrating \eqref{conseq} over $M$ leads to the condition (remember $\left.\chi^a\right|_{\partial M}=\zeta^a$, $\left.\chi^\perp\right|_{\partial M}=0$):
   \begin{equation}
   0=\int_{\partial M} \sqrt{k}d^2y\, j^\perp=\langle \zeta,\zeta \rangle_h \, .\label{nullcond}
   \end{equation}
   We thus find that the equation \eqref{conseq} implies that the boundary vector field $\zeta$ is a null vector for the metric \eqref{bndmetric2}. Note that we showed in \eqref{mixed} that \eqref{bndmetric2} has mixed signature and hence allows null vectors. So we see that \eqref{nullcond} provides an integrability condition on the boundary values of $\chi$ for which the problem (\ref{const1vec}, \ref{const2vec}) has a solution. From a physical point of view we can consider this integrability condition as a constraint on the geodesic problem \eqref{action2}, selecting the null geodesics. 
   
   In summary consideration of the Hamiltonian constraint solves both puzzles listed at the end of the previous subsection (albeit the second only conjecturally): 1) only metrics corresponding to null geodesics solve the Einstein equations including the Hamiltonian constraint 2) given a null boundary vector field there exists a unique bulk extension satisfying both momentum and Hamiltonian constraints.

   \subsection{The homogeneous space structure}\label{homspac}
   It followed rather directly from the definition of the set of vacuum metrics that there is a transitive action of diffeomorphisms on it. At the same time many -- though not all -- of these metric should be physically identified as describing the same space-time. We will now shortly review how the analysis of the constraints is equivalent with the quotient \eqref{homspac1} and how the metric \eqref{bndmetric1} is consistent with the homogeneous space structure. For an analogous -- but not completely equivalent -- discussion in more detail we refer to \cite{Seraj:2017rzw}.
   
   \subsubsection*{The first quotient via the constraints}
   It might so far have appeared somewhat arbitrary to only physically identify metrics that differ by diffeomorphisms that act trivially on the boundary. Let us explain here how this is reflected in the constraints. In the previous subsections we investigated the tangent space of the space of vacua by considering motion along it. The momentum constraint equations (\eqref{const1}, \eqref{const2}) on the velocities \eqref{tangent vectors} can as such be interpreted as constraints on the vectors tangent to $\calv$. As we partially argued, partially conjectured there, every boundary vector field that has vanishing norm with respect to the inner product \eqref{bndmetric2} has a unique bulk extension consistent with the constraints. This meas that we can uniquely split any bulk vector field  $\chi$ with null boundary component $\zeta$ as:
   \begin{equation}
   \chi=\chi[\zeta]+\chi_0 \, .
   \end{equation} 
   Here $\chi[\zeta]$ is the vector field obtained by restricting $\chi$ to the boundary and then solving the constraints with these boundary values as input. The other part $\chi_0$ is by definition simply the difference $\chi-\chi[\zeta]$. The point is now that by definition $\chi_0$ is a vector field that vanishes on the boundary, or in other words it is an element of the Lie algebra $\calsg_0$ of the group $\calg_0$ of boundary trivial diffeomorphisms. One easily verifies that $\calsg_0$ forms a Lie-algebra ideal of the Lie algebra $\calsg$ of boundary preserving diffeomorphisms. We can thus think of the $\chi[\zeta]$ as representatives for the quotient Lie algebra $\cals=\calsg/\calsg_0$, and it is exactly these that are selected by the constraints. So an analysis of the constraints implies one should quotient the group of boundary preserving diffeomorpisms by the normal subgroup of diffeomorphisms that vanish at the boundary. By the first isomorphism theorem the resulting quotient group is isomorphic to the group of boundary diffeomorphisms:
   \begin{equation}
   \cals\equiv\frac{\calg}{\calg_0}\cong \mathrm{Diff}(\partial M)\, .
   \end{equation}
   Note that strictly speaking we skipped a few logical steps in the above argument; we started our argument by considering bulk vector fields whose boundary components were null with respect to \eqref{bndmetric2} and ended up drawing conclusions for all bulk vector fields. That indeed such a generalization is natural and valid is supported by the fact the the metric \eqref{bndmetric2} is well defined on (a further quotient of) $\cals$ and furthermore invariant under its left-action, as we will explain below. The main conclusion is that we should interpret $\phi$ in \eqref{vacdef} as an element of $\cals$.

   \subsubsection*{Left invariant metric and a second quotient}
   One of the main results of the previous subsection was the derivation of the inner product $\langle \zeta,\zeta \rangle_{\bar h}$ \eqref{bndmetric2} on the space of vacua. It is important to note that this metric depends on the vacuum $\bar h$. This dependence is however highly restricted, due to the homogeneous space structure of the space of vacua. Due to the transitive action of $\cals$ the metric at an arbitrary point $\bar h$ can be easily related to the metric at a fixed reference point $\bar h_o$. This general feature of homogeneous spaces can be made fully explicit in our setup as well by re-expressing a tangent vector \eqref{tangent vectors} at $\bar h$ in terms of one at $\bar h_o$ (see appendix \ref{techtime} for a derivation):
    \begin{equation}\label{tangent vectors sigma}
      \dot h=\phi_z\cdot L_{\sigma_z}\bar h_o=\phi \cdot D_o\sigma_z \quad \mbox{where} \quad \sigma_z^i(x)=\dot \phi_z^i (\phi^{-1}_z(x))\, .
    \end{equation}
    In terms of vector fields this amounts to $\chi_z=\phi_z\cdot\sigma_z$.  By diffeomorphism invariance of \eqref{bndmetric1}/\eqref{bndmetric2} it then follows that (identifying bulk vector fields with their boundary values, as explained in the previous subsections):
    \begin{equation}
    \langle\chi_z,\chi_z\rangle_{\bar h(z)}=\langle\sigma_z,\sigma_z\rangle_{\phi^{-1}\cdot\bar h(z)}=\langle\sigma_z,\sigma_z\rangle_{\bar h_o}\equiv\langle\sigma_z,\sigma_z\rangle \, .
    \end{equation}
    The key point of the line above is that all the way on the left we have a $z$-dependent inner product of two $z$-dependent vectors, while all the way on the right we have a $z$-independent inner product -- where for this reason we drop the $\bar h_0$ subscript -- of a $z$-dependent vector field. This is the key property of a homogeneous space, and $\sigma_z$ plays the role of local frame on $\calv$. Note that more pragmatically $\sigma_z$ is a boundary vector field and $\langle\cdot,\cdot\rangle$ hence an inner product on the Lie algebra of boundary diffeomorphisms. It is useful to compare it to the standard\footnote{See e.g. \cite{arnold66}.} inner product:
    \begin{equation}
    (\zeta_{(1)},\zeta_{(2)})=\int_{\partial M}\sqrt{k_o}d^2y\, k_{ab}^o\zeta_{(1)}^a\zeta_{(2)}^b \, .\label{bilin}
    \end{equation}
    Via \eqref{bndmetric2} one then has the relation
    \begin{equation}
    \langle \zeta_{(1)},\zeta_{(2)}\rangle=(\zeta_{(1)},\mathbb{D}\zeta_{(2)})\label{Dmetric}
    \end{equation}
    where
    \begin{equation}
    \mathbb{D}\zeta^a=D^\perp \zeta^a-K^a{}_b\zeta^b \,.\label{DDdef}
    \end{equation}
   This operator $\mathbb{D}$ that determines the inner product has various properties. Note first that $\mathbb{D}$ is symmetric with respect to \eqref{bilin}, as can be established via the results of the previous subsections. Second we observe that any boundary preserving bulk Killing vector is a boundary Killing vector and sits in the kernel of $\mathbb{D}$, this follows directly from decomposing the bulk Killing equation on the boundary via \eqref{cddecomp} together with (\ref{boundcond}, \ref{bcond2}):
   \begin{equation}
   \nabla_{(i}\chi_{j)}=0\quad\Rightarrow\quad \left.D^\perp\chi^\perp\right|_{\partial M}=0\,,\ \ D_{(a}\zeta_{b)}=0\,,\ \ D^\perp \zeta^a-K^a{}_b\zeta^b=0 \, .\label{boundkill}
   \end{equation}
   One should then note that the Killing vectors are exactly those that leave the vacua invariant and hence form the isotropy group $\calk$ of the transitive action of diffeomorphisms on the space of vacua. Putting these observations together establishes that \eqref{Dmetric} is a metric on the homogeneous space $\calv=\cals/\calk$, the quotient of the space of boundary diffeomorphisms by boundary isometries.  It also follows that $\mathbb{D}$ is $\Ad(\calk)$-equivariant, making \eqref{Dmetric} left invariant under the action of $\cals$. In summary $\calv$ together with the metric \eqref{Dmetric} is a (infinite dimensional) Riemannian homogeneous space\footnote{See appendix B of \cite{Seraj:2017rzw} for our definitions and conventions on homogeneous spaces.}.
   
   An interesting question that remains is if the kernel of $\mathbb{D}$ is also spanned by Killing vectors. If not then the metric \eqref{Dmetric} would be degenerate. Although we expect \eqref{Dmetric} to be non-degenerate on $\calv$ we leave a detailed argument to future work. In the example of a round boundary indeed this expectation is true, see section \ref{spexample}.
   
   The homogeneous space structure is very relevant also for the geodesic problem, as (most) geodesics are simply orbits of one-parameter subgroups of $\cals$. We will not investigate this further here but refer the interested reader to \cite{Seraj:2017rzw} where this was worked out in more detail for the case of YM theory.
   
   \paragraph{Asymptotic symmetries as conserved momenta}
   The group $\cals$ discussed above is that of symmetries of the adiabatic action \eqref{action2} and thus lead to conserved momenta through the standard Noether procedure. The momentum associated to a boundary diffeomorphism generated by a vector field $\zeta$ is given through the Noether formula by
   \begin{equation}\label{momenta GR}
   P_\zeta\equiv\rg_z(\dot z,\de_\zeta h)=\langle\zeta,\chi_z\rangle_{\bar h(z)}=\int_{\partial M}\sqrt{\bar k}d^2y\,n^ie^j_a\zeta^a\dot h_{ij} \,.
   \end{equation}
   These conserved momenta coincide with the conserved charges of full GR in our setup\footnote{The same is true in YM theory, see \cite{Seraj:2017rzw}.}. To make this explicit, we will show that exactly the same result \eqref{momenta GR} is found by a symmetry analysis through the covariant phase space method (see e.g. \cite{Compere:2018aar} for an overview). First one defines the symplectic potential $\Theta(\de h)=\int \theta$,  through the total derivative term appearing in the variation of the Lagrangian $\de \cL=(\mathrm{E.o.m})\de h+d\theta(\de h)$ and the integration is over a Cauchy hypersurface -- $M$ in our particular case. Then the symplectic form of the theory is defined as the antisymmetric variation of the symplectic potential $\Omega(\de_1,\de_2)=\de_1\Theta(\de_2)-\de_2\Theta(\de_1)$.  Now the charges are defined by contracting the associated symmetry with the symplectic form i.e.  \begin{align}
   \de Q_\chi&=\Omega (\de h, \de_{\chi}h)
   \end{align}
   where in this case $\de_\chi=\cL_\chi$ is the Lie derivative with respect to a  diffeomorphism $\chi$. Hence
   \begin{align}
   \de_{\chi}\Theta(\de h)&=L_\chi\Theta(\de h)=\chi\cdot d\Theta+d(\chi\cdot\Theta) \, .
   \end{align}
   Integrating this over $M$ and using that $\chi$ is a spatial vector with vanishing normal component at the boundary leads to 
   \begin{align}
   \de Q_\chi&=\int_M\de \Theta(\de_{\chi}h)
   \end{align}
   which can be directly integrated to give
   \begin{align}
   Q_\chi&=\int_M \Theta(\de_{\chi}h)\,.
   \end{align}
   The symplectic potential $\Theta$ of GR -- see \cite{Lee:1990nz} -- in the GNC gauge \eqref{mGNC}, evaluated on a diffeomorphism $\delta h_{ij}=2\nabla_{(i}\chi_{j)}$ is given by 
   \begin{align}
   \Theta(\de_\chi h)=\int_M d^3x\; \sqrt{h}(\dot h_{ij}-h_{ij}h^{kl}\dot h_{kl}) \nabla^i\chi^j \, .
   \end{align}
   Through integration by part and using the momentum constraint \eqref{const1} one finds
   \begin{align}\label{charge}
   Q_\chi&=\int_{\pM}d^2y \sqrt{k} \,n^i\, \chi^j \dot h_{ij} \, ,
   \end{align}
  which confirms
  \begin{equation}
  P_\zeta=Q_\chi\quad\mbox{where}\quad \zeta^a=\left.\chi^a\right|_{\partial M} \, .
  \end{equation}

\section{Example: GR in a round sphere} \label{spexample}
In this section we illustrate some parts of our construction by working it out in the simplest case, where the boundary is a round 2-sphere, in other words we consider $M=B^3_R$ the three dimensional ball of radius $R$ with its standard embedding in $\mathbb{R}^3$. We will choose as our reference coordinates the spherical ones, $(x^i)=(r,\theta,\varphi)$, so that the boundary is simply the surface $r=R$ and the reference vacuum can be chosen to be
\begin{equation}
ds^2_o=h_{ij}^{o}dx^idx^j=dr^2+r^2(d\theta^2+\sin^2\theta d\varphi^2)\, .
\end{equation}
The natural foliation to use is then by spheres of radius $r$ so that
\begin{equation}
n_i=\delta_i^r \, ,\qquad e_a^r=0 \, ,\qquad e^a_b=\delta^a_b \, ,\qquad ds^2_*=k_{ab}dy^ady^b=r^2(d\theta^2+\sin^2\theta d\varphi^2)\, .
\end{equation}
Using these expressions one finds $\Gamma_a^b=K_a{}^b=\frac{1}{r}\delta^a_b$ so that
\begin{equation}
\mathbb{D}\zeta^a=\partial_r\zeta^a
\end{equation}
and thus
\begin{equation}
\langle\zeta_{(1)},\zeta_{(2)}\rangle=\oint_{S^2_R}d^2 y\,\sqrt{k} k_{ab}\zeta^a_{(1)}\partial_r\zeta_{(2)}^b \, .
\end{equation}
The key point is of course the meaning of the radial derivative $\partial_r$ on the boundary vector field $\zeta$. As explained in section \ref{momsec}, it is defined as the radial derivative of the bulk extension of this boundary vector field. Instead of relying on the general results used in that section we can find the extension in this case by direct construction. Using the Helmholtz split \eqref{chidecomp}, $\chi_i=\eta_i+\partial_i\phi$,  one can solve the momentum constraint \eqref{const1vec} for $\eta$. The unique solution with vanishing normal component at the boundary is
\begin{equation}
\eta^i=r\sqrt{h}\epsilon^{ijk}n_j\partial_k H
\end{equation}
where $H$ is a harmonic function:
\begin{equation}
H=-R\sum_{\ell,m} a_{\ell m} \left(\frac{r}{R}\right)^\ell Y_{\ell m} \, .
\end{equation}
It follows that on the boundary
\begin{equation}
\left.\eta^a\right|_{\partial M}=R^2 \sqrt{k}\epsilon^{ab}\partial_b \tau\,, \qquad\quad \tau=\sum_{\ell,m} a_{\ell m} Y_{\ell m}\,.
\end{equation}
We thus see that the split \eqref{split} of the boundary vector fields in this case amounts to the Hodge decomposition on the sphere:
\begin{equation}
\zeta_a=\left.\chi_a\right|_{\pM}=\left.\eta_a\right|_{\partial M}+\partial_a\sigma=R^2 (\sqrt{k}\epsilon_a{}^{b}\partial_b \tau+\partial_a\rho) \, , \quad \quad \rho=\sum_{\ell,m} b_{\ell m} Y_{\ell m} \, . \label{decompsp}
\end{equation}
Given this decomposition we can then compute the normal derivative via the normal derivative for the bulk extension:
\begin{eqnarray}
\mathbb{D}\zeta^a=\partial_r\zeta^a&=&\left.(\partial_r\eta^a)\right|_{\partial M}+\left.(k^{ab}\partial_a\partial_r\phi)\right|_{\partial M}+\left.(\partial_a\phi\partial_r k^{ab})\right|_{\partial M}\\
&=&R \left[\sqrt{k}\epsilon^{ab}\partial_b\left(\sum_{\ell,m} (\ell-1) a_{\ell,m} Y_{\ell m}\right)-2k^{ab}\partial_b\rho\right]\label{dsphere}.
\end{eqnarray}
Note that although the extension in the bulk of $\phi$ is not fully determined by the momentum constraint the boundary condition that $\partial_r\phi$ vanishes all over the boundary is enough to perform the above computation. This holds more generally, see formula \eqref{phider} with which the result above is consistent. In principle one can obtain the unique bulk extension $\partial_i\phi$ of $\partial_a\sigma$ by the Hamiltonian constraint \eqref{const2vec}, but we have so far failed to solve this equation explicitly, even in the particular example considered here.

Given the explicit expression \eqref{dsphere} one can directly read of the kernel, it is given by the divergence free vectors for which $\tau$ has only an $\ell=1$ component. These are exactly the vector fields that generate the SO(3) rotations, confirming our expectation that in general this kernel is given by boundary preserving Killing vectors. We can then fully characterize the space of vacua in this example:
\begin{equation}
\calv=\mathrm{Diff}(S^2)/\mathrm{SO}(3) \, .
\end{equation}
Furthermore via \eqref{dsphere} we can also characterize the metric at the reference point explicitly:
\begin{equation}
\langle\zeta_{(1)},\zeta_{(2)}\rangle=R^3\sum_{\ell,m} \ell (\ell+1)\left( (\ell-1) a^{(1)}_{\ell m} a^{(2)}_{\ell m} - 2 b^{(1)}_{\ell m} b^{(2)}_{\ell m}\right) \, .\label{spmetric}
\end{equation}
Note that in this particular example  the decomposition \eqref{decompsp} is also orthogonal. The positive definite component is spanned by volume preserving diffeomorphisms.

\section{Discussion}\label{discs}
In this paper --  as a continuation of \cite{Seraj:2017rzw} -- we investigated a subset of the configuration space of GR that we call the space of vacua. In the presence of a boundary for the spatial manifold, those diffeomorphisms that are non-trivial on the boundary generate global symmetries rather than a gauge equivalence. These symmetries are spontaneously broken and generate the whole space of vacua so that it is a homogeneous space. One consequence of the presence of this space of vacua is that there will exist in GR -- in the presence of a finite boundary -- slow velocity solutions that can be approximately described as geodesic motion on this space. We computed the corresponding metric \eqref{bndmetric2} on the space of vacua, showed how it is fully determined in terms of data on the boundary and computed its pseudo-Riemannian signature. Furthermore we explained how the Hamiltonian constraint becomes a constraint on the geodesic motion on the space of vacua, selecting only null geodesics. 

There remain a few open issues in our analysis. The main problem that requires further investigation is our conjecture of uniqueness and existence of solutions to the boundary value problem \eqref{phieq}. Finding (example) solutions would allow for the explicit construction of nontrivial approximate solutions of the 4d vacuum Einstein equations, which would of course be highly interesting.

Furthermore there are some obvious generalizations of our setup to consider. One could investigate more general topologies for $M$ and $\partial M$ or repeat a similar analysis around other solutions, such as black holes.

As far as we are aware our approach has not been considered previously in the literature (apart from \cite{Lechtenfeld:2015uka} for YM). Although our starting point and method is new in its application to GR, there are various concepts and results that are closely related to various other areas of recent research. Below, we discuss a few of them that deserve further investigation.

\subsection*{Relation to edge modes} 
One interpretation of our results is that the dynamics of GR in an adiabatic limit is described in terms of boundary degrees of freedom only. This is still far from full-fledged holography, which tries to reformulate all of GR in terms of boundary degrees of freedom. It appears our adiabatic solutions are closely related to edge modes --  degrees of freedom due to the presence of a boundary in addition to bulk degrees of freedom. To keep the discussion short, we consider the electromagnetic couterpart of our analysis. The presymplectic potential \cite{Lee:1990nz} of the theory is given by
$\Theta=\int_M d^3x\, E^i \,\de A_i$  which has the structure $p^i\de q_i$ pairing canonical variables. To extract the bulk and boundary degrees of freedom we use the Hodge decomposition (see equation \eqref{excc}) to expand $E^i, A_i$ into their transverse and longitudinal parts
\begin{align}\label{E-decomposed}
E_i&=\hat{E}_i+\pd_i \varphi \, ,\qquad \pd_i \hat E^i=0 \, ,\qquad \hat E\cdot n=0\, ,\qquad \cd^2\varphi=0 \, ,\\
A_i&=\hat{A}_i+\pd_i \psi \, ,\qquad \pd_i \hat A^i=0 \, ,\qquad \hat A\cdot n=0 \, .
\end{align}
The last equality in \eqref{E-decomposed} is a result of $\pd_i E^i=\pd_i \hat E^i=0$. Using the above decompositions in the symplectic potential we find
\begin{align}
\nonumber\Theta&=\int_M d^3x\, \hat{E}_i\, \de\hat{A}_i +\oint_{\pd M}\pd_\perp \varphi\, \de \psi=(\hat{E},\de \hat A)+\langle \varphi,\de\psi\rangle \, , 
\end{align}
where the inner products are those of the configuration space and the space of vacua  (the electromagnetic couterparts of \eqref{WdW}, \eqref{bndmetric1}) respectively.
Thus we see that while the transverse parts of the electric field and gauge field pair in the bulk, the normal component of the electric field is paired with a large gauge field at the boundary. The latter pair is referred to as a boundary degree of freedom, or edge mode, and has appeared in a number of different analyses, see e.g. \cite{Strominger:2017zoo,Barnich:2019xhd,Donnelly:2016auv,Blommaert:2018oue}. Our analysis however gives a bulk interpretation to the latter as well. It pairs the Coulombic part of the electric field with the harmonic extension of large gauge field inside the bulk. The next question is to find the theory governing the dynamics of edge modes. This question has been tackled in \cite{Barnich:2019xhd,Blommaert:2018oue}. Our adiabatic analysis induces a dynamics for edge modes which at first glance is a truncation of that of \cite{Blommaert:2018oue}\footnote{We thank A. Blommaert and T. Mertens for an interesting discussion of this point.}, while the connection to the boundary action of \cite{Barnich:2019xhd} is less clear. Clarifying these issues would be interesting.

\subsection*{Infinite volume limit and relation to soft charges and IR structure}
In the last few years there has been a renewed interest in asymptotic symmetries, with the establishment of a direct connection to soft theorems and the memory effect -- see \cite{Strominger:2017zoo} for a review and a stepping stone into the literature. By its very definition this subject deals with infinite volume while in our approach we consider a compact spatial manifold. It is however natural to investigate whether the results in infinite volume can be reproduced as a limit of our analysis. In YM theory the symmetry groups were identical for infinite volume and finite volume, while this matching is not obvious for GR where the asymptotic symmetries contain the BMS group. Our result could possibly be matched with an extension of the BMS group containing Diff$(S^2)$. One such extension was proposed in \cite{Campiglia:2014yka}, but especially the symmetries of \cite{Hamada:2018vrw} (see also \cite{Mirbabayi:2016xvc}) -- obtained by a correspondence to an infinite tower of soft theorems -- appear to be identical to those of our setup.

In the limit of infinite volume at finite time, the boundary will tend to spatial infinity and recently the asymptotic symmetries of GR have been reformulated in that setting \cite{Henneaux:2018cst, Henneaux:2018hdj, Henneaux:2019yax,Compere:2017knf}. Let us mention that understanding asymptotic symmetries through a limit from finite volume has been previously considered in \cite{Andrade:2015fna, Riello:2019tad}. 

One of the main attractions of our approach is its highly geometric nature. Recently an equally geometric approach to asymptotic/boundary symmetries has been developed \cite{Gomes:2018shn, Gomes:2018dxs, Riello:2019tad}. There -- using the natural metric on configuration space -- a connection which singles out the transverse (radiative) degrees of freedom in phase space is defined. Note however that this does not automatically imply that the longitudinal sector is non-physical. Indeed we showed here and in \cite{Seraj:2017rzw} that longitudinal dynamics can lead to physically non-trivial solutions, which find themselves naturally in the Coulombic sector of the theory (see also \cite{Barnich:2010bu, Barnich:2019xhd}).

Finally let us point out the similarity of our starting point -- namely a family of metrics related by diffeomorphisms, see \eqref{intro-timedep} --  to that of Weinberg in his work on adiabatic modes in cosmology \cite{Weinberg:2003sw}. One apparent difference is however that there one starts from a limit of small spatial derivatives while we consider small time derivatives. It would be interesting to see if the geometric methods we apply can play a role in this area of research -- see e.g. \cite{Hui:2018cag,Pajer:2017hmb} for recent developments.

\subsection*{Relation to multipoles}

In YM theory we observed -- by explicit construction \cite{Seraj:2017rzw} -- that various static electric fields which are manifestly non-pure gauge are produced by time-dependent spatial gauge transformations of the trivial gauge field. Although we have not been able to explicitly construct the corresponding solutions in GR, our work shows that at least implicitly the same is true in GR. The conserved charges of the geodesic motion in YM theory have a direct interpretation as multipole charges \cite{Seraj:2016jxi, Seraj:2017rzw}. It seems natural to expect the same to be true in the case of GR, especially given the recent reformulation of gravitational multipole moments as Noether charges \cite{Compere:2017wrj}. A technical difficulty in such a comparison is however the different choice of gauge made in these two approaches.

\section*{Acknowledgements}
We thank A. Blommaert, T. Mertens and B. Oblak for interesting and useful discussions. ESK and DVdB are partially supported by TUBITAK grant 117F376 and DVdB is also partially supported by the Bo\u{g}azi\c{c}i University Research Fund under grant number 17B03P1. AS is supported by a scientific collaborator grant of the Fund for Scientific Research-FNRS Belgium.
AS was supported by a TUBITAK 2221 grant during a visit to Bo\u{g}azi\c{c}i University while part of this work was undertaken.

\appendix

\section{Technicalities on the time derivative of a spatial diffeomorphism}\label{techtime}
It will be convenient to use the shorthand notation
\begin{equation}
\phi^{\uk}_i=\partial_i\phi^k \, , \qquad \phi_i^\uk\phi^j_\uk=\delta_i^j \, ,\qquad \phi_i^\uk\phi^i_\ul=\delta_\ul^\uk \, .
\end{equation}
Our starting point are the formulae \eqref{difdef} and \eqref{vacdef} that can be combined together as
\begin{equation}
h_{ij}(x)=\phi^\uk_i\phi^\ul_j h^o_{kl}(\phi(x)) \, .
\end{equation}
We then compute the time derivative, using that time only enters through $\phi$:
\begin{equation}
\dot h_{ij}(x)=\dot\phi^\uk_i\phi^\ul_j h^o_{kl}(\phi(x))+\phi^\uk_i\dot\phi^\ul_j h^o_{kl}(\phi(x))+\phi^\uk_i\phi^\ul_j \dot\phi^{m}\partial_m h^o_{kl}(\phi(x)) \, .\label{timetemp}
\end{equation}
There are now two ways to rewrite this. First observe that
\begin{eqnarray}
L_\chi h_{ij}&=&\chi^m\partial_m h_{ij}+2h_{m(i}\partial_{j)}\chi^m\\
&=&2\chi^m \phi_{m(i}^\uk\phi^\ul_{j)}h^o_{kl}+\phi_{i}^\uk\phi_j^\ul\chi^m \phi_{\underline m}^n\partial_nh^o_{kl}+2h_{kl}^o\phi^\uk_m\phi^\ul_{(i}\partial_{j)}\chi^m\\
&=&\phi_{i}^\uk\phi_j^\ul\chi^m \phi_{\underline m}^n\partial_nh^o_{kl}+2h_{kl}^o\phi^\ul_{(i}\partial_{j)}\left(\phi^\uk_m\chi^m\right) \, .
\end{eqnarray}
So we see that 
\begin{equation}
\dot h_{ij}(x)=L_{\chi_z}h_{ij}(x)\, ,\qquad \chi_z^i=\dot\phi^k \phi_\uk^i \, .
\end{equation}
This first way of rewriting expresses the intuitive fact that the time derivative (i.e. velocity) at some point is given by a tangent vector at that point (and furthermore that this tangent vector is the image of a tangent vector on the group manifold). There is also a second way of looking at things, which will be more useful for us, by mapping this tangent vector to the fixed reference point $h^o$. To see this we rewrite \eqref{timetemp} in a second way:
\begin{equation}
\phi\cdot(L_\sigma h^o)_{ij}(x)=\phi_i^\uk\phi_j^\ul\left(\sigma^m(\phi(x))\partial_m h_{kl}^o(\phi(x))+2h_{m(k}^o(\phi(x))\partial_{l)}\sigma^m(\phi(x))\right)\, .
\end{equation}
We thus see that we reproduce \eqref{timetemp} if we take
\begin{equation}
\sigma^m(x)=\dot\phi^m(\phi^{-1}(x)) \, .
\end{equation} 
This follows because $\partial_l\sigma^k=\dot\phi^{\uk}_i\phi^i_\ul$. So we get
\begin{equation}
\dot h_{ij}(x)=\phi\cdot(L_{\sigma_z} h^o)_{ij}(x)\,.
\end{equation}
Finally observe that
\begin{equation}
\chi_z=\phi\cdot \sigma_z \, .
\end{equation}
The two relations can then also be seen to follow from the identity
\begin{equation}
L_{\phi\cdot\sigma}(\phi\cdot h_o)=\phi\cdot\left(L_\sigma h_o\right) \, .
\end{equation}

\section{Details of the boundary geometry}\label{fol}
In the main text we use various geometric aspects -- both intrinsic and extrinsic -- of the boundary, that follow naturally from the embedding of the boundary manifold $\partial M$ into the manifold with boundary  $M$. The collar theorem \cite{Schwarz:1995} guarantees that the boundary can be described as a hypersurface in a foliation of a neighborhood of the boundary in $M$. We can thus express the boundary geometry using the language of foliations, see e.g \cite{Gourgoulhon:2007ue}.We will use the indices $i,j,k,\ldots$ and $a,b,c,\ldots$ when referring to the tangent space of $M$ and $\partial M$ respectively. Note that $M$ is the spatial manifold of space-time and has a metric $h$ of Euclidean signature, as does the induced metric $k$ on the boundary, and we raise/lower indices with these metrics respectively.

We start by introducing a normalized outward pointing normal vector  $n^i$ and $d-1$ vectors $e_a^i$, where $a:1,...,d-1$, that will be tangent to the hypersurfaces:
\begin{equation}
n_in^i=1 \, , \qquad e^i_a n_i=0 \, .
\end{equation}
A concrete realization of these vectors can be made when the hypersurfaces in the foliation are defined by a particular scalar function being constant: $F(x)=\mathrm{cst}$. There then exist $d-1$ parameters $y^a$ -- coordinates on the boundary -- and functions $x^i(y)$ such that $\partial_a F=0$. In that case one can choose
\begin{equation}
e^i_a=\frac{\partial x^i}{\partial y^a} \, ,\qquad  n_i=N\partial_{i}F\quad \mbox{where} \quad N=(h^{ij}\partial_iF\partial_jF)^{-1/2} \, .
\end{equation} 
The bulk metric $h$ then decomposes in terms of the induced metric $k$ as follows
\begin{equation}
h^{ij}=n^i n^j+e_a^ie_b^j k^{ab}\qquad \mbox{where}\qquad k_{ab}=h_{ij}e_a^ie_b^j\,.
\end{equation}
Furthermore one defines
\begin{equation}
e_i^a=h_{ij}k^{ab}e_b^j \, .
\end{equation}
The boundary geometry, or more concretely its metric and all bulk covariant derivatives restricted to the boundary, can be expressed in terms of a few basic objects: the induced metric $k_{ab}$, the extrinsic curvature $K_{ab}$, the `acceleration vector' $\kappa_a$, a `normal connection' $\Gamma_a^b$ and finally the Levi-Civita connection of the induced metric $\Gamma_{ab}^c$. These objects are defined as
\begin{eqnarray}
K_{ab}&=&e_a^ie_b^j \nabla_i n_j \, ,\\
\kappa_a&=&n^ie^j_a\nabla_in_j \, ,\\
\Gamma_a^b&=&n^i e^b_j\nabla_i e^j_a \, ,\\
\Gamma_{ab}^c&=&e^c_ie_b^j\nabla_j e_a^i
\end{eqnarray}
where $\nabla$ is the Levi-Civita connection for $h$. For the special cases 1) $ \left[ n, e_a \right]=0$ one has $\Gamma_a^b=K_a^b$, 2) $n$ generates $M$-geodesics, $\kappa_a=0$.  For calculational purposes the decomposition of the covariant derivatives of the foliation adapted frame in terms of the above objects is useful:
\begin{eqnarray}
\nabla_i n_j&=&\kappa_a e^a_j n_i + K_{ab}e_i^ae_j^b \, ,\\
\nabla_i n^j&=&\kappa^a e_a^j n_i + K_{a}{}^be_i^ae_b^j\, ,\\
\nabla_i e^j_a&=&-\kappa_an_in^j-K_{ab}n^je_i^b+\Gamma_a^b n_ie_b^j+\Gamma_{ab}^c e^b_ie_c^j \, ,\\
\nabla_i e_j^a&=&-\kappa^an_in_j-K^{a}{}_b n_je^b_i-\Gamma_b^a n_ie^b_j-\Gamma_{bc}^a e^b_ie^c_j \, .
\end{eqnarray}
A bulk vector can be decomposed into parts normal and tangent to the hypersurfaces as
\begin{equation}
V^i=V^\perp n^i+e^i_a V^a
\end{equation}
and the bulk covariant derivative of such a vector then decomposes as
\begin{eqnarray}
\nabla_i V^j&=&\left(D^\perp V^\perp-\kappa_a V^a\right)n_in^j+\left(D_a V^\perp-K_{ab} V^b\right)e_i^an^j\nonumber\\
&&+\left(D^\perp V^a+\kappa^aV^\perp\right)n_ie^j_a+\left(D_a V^b+K_{a}{}^bV^\perp\right)e_i^ae^j_b \label{cddecomp}
\end{eqnarray} 
where
\begin{eqnarray}
D^\perp V^\perp&=&n^i\partial_i V^\perp \, ,\\
D^\perp V^a&=&n^i\partial_i V^a+\Gamma^a_bV^b \, ,\\
D_a V^\perp&=&\partial_a V^\perp \label{ndhyptang} \, ,\\
D_a V^b&=&\partial_a V^b+\Gamma_{ac}^bV^c \, .
\end{eqnarray}
Note that the hypersurface covariant derivatives $D^\perp$ and $D_a$ naturally generalize to arbitrary hypersurface tangent tensors and both are metric compatible
\begin{equation}
D^\perp k_{ab}=0 \, ,\qquad D_a k_{bc}=0 \, .
\end{equation}

\section{The boundary value problem via differential form decomposition}\label{apphodge}
In the main text of this paper an important boundary value problem appears. Given a tangent vector field, what are the possible bulk extensions with vanishing normal component at the boundary that solve the momentum constraint \eqref{const1vec}? We shortly discussed the solution of this problem there, but will fill in some details and make things more precise in this appendix. The solution of the problem is based on the Helmholtz decomposition, namely that every vector field can be written as a sum of a gradient and curl. This classic theorem is more elegantly phrased in the modern language of differential forms on manifolds with boundaries where its generalization is the Hodge-Morrey-Friedrichs decomposition. This appendix is based on the introduction to the subject \cite{Schwarz:1995} whose notation we follow. We start by reviewing some key notation and results and then use these to address the particular boundary value problem of interest in this paper. 

\subsection{Hodge-Morrey-Friedrichs decomposition}
The starting point is a compact Riemannian manifold with boundary\footnote{Strictly speaking $M$ is assumed to be a $\partial$-manifold, which demands additionally topological completeness, differentiability and orientation \cite{Schwarz:1995}.} $(M, h)$, where we denote the boundary by $\partial M$, for a full definition see \cite{Schwarz:1995}. We will be dealing with differential forms $\omega\in \Lambda^p(M)$, on which we have the usual definitions of exterior derivative $d$, hodge star $\star$ and co-differential $\delta=(-1)^{p(d-p+1)}\star d\star$. 

Any differential form $\omega$ can be restricted to the boundary, and we denote this restriction with $\left.\omega\right|_{\partial M}$. This should not be confused with the pull-back -- via the natural inclusion map $\imath : \partial M\rightarrow M$ -- of $\omega$ to the boundary, which has only components tangent to the boundary and which for this reason we denote with $\mathrm{t} \omega$. The normal part of a form on the boundary is then defined to be
\begin{equation}
\mathrm{n}\omega=\left.\omega\right|_{\partial M}-\mathrm{t} \omega \, .
\end{equation}
It follows from these definitions that
\begin{equation}
d\rt\omega=\rt d\omega\qquad \mbox{and} \quad \delta\rn\omega=\rn\delta\omega \label{boundrels}
\end{equation}
where with slight abuse of notation the exterior derivative/co-differential on $M$ and $\partial M$ are indicated with the same symbol. As in the case without boundary $\Lambda^k(M)$ comes equipped with the inner product
\begin{equation}
\langle \omega,\eta\rangle=\int_M \omega\wedge\star\eta \, .\label{inner}
\end{equation}
This inner product can then be used to define the Hilbert space of square integrable $k$-forms and in this paper we only consider such square integrable forms.

The first non-trivial consequence of a non-empty boundary is that $d$ and $\delta$ are no longer automatically each others adjoint, but that this depends on boundary conditions. This follows from the Green's identity 
\begin{equation}
\langle d\omega,\eta\rangle=\langle \omega,\delta \eta\rangle+\int_\pM \rt\omega\wedge\star\rn\eta \, .\label{green}
\end{equation}
This formula is one of the main technical tools underlying the derivations of the following results.
\paragraph{Hodge-Morrey decomposition} (\cite{Schwarz:1995} Thm 2.4.2)
One can {\it uniquely} decompose a $k$-form $\omega$ on $M$ as
\begin{equation}
\omega=d\alpha+\delta\beta+\kappa\qquad\mbox{where}\quad \rt\alpha=0\,,\quad \rn\beta=0\quad\mbox{and}\quad d\kappa=\delta\kappa=0 \, . \label{hodgemorrey}
\end{equation}
Note that this decomposition is orthogonal, i.e. the three terms on the RHS are each mutually orthogonal with respect to the inner product \eqref{inner}. A form like $\kappa$, that is both closed and co-closed is called a {\it harmonic field}\footnote{Note that every harmonic field is automatically a {\it harmonic form}, i.e. a form on which the Laplace-Beltrami operator $\Delta=d\delta+\delta d$ vanishes. On a manifold with boundary the reverse is however not always true, not every harmonic form is a harmonic field.}.
\paragraph{Hodge-Morrey-Friedrichs decomposition} (\cite{Schwarz:1995} Cor 2.4.9)
The decomposition \eqref{hodgemorrey} can be further refined by additionally decomposing the harmonic field $\kappa$. This can be done in 2 ways, each providing a unique decomposition of an arbitrary $k$-form $\omega$:
\begin{eqnarray}
\omega&=&d\alpha+\delta\beta+d\lambda+\check\kappa\quad\mbox{where}\quad \rt\alpha=\rn\beta=\rn\check\kappa=d\tilde\kappa=\delta\check\kappa=\delta d\lambda=0 \, , \label{hmf1}\\
\omega&=&d\alpha+\delta\beta+\delta\gamma+\hat\kappa\quad\mbox{where}\quad \rt\alpha=\rn\beta=\rt\hat\kappa=d\hat\kappa=\delta\hat\kappa=d\delta\gamma=0 \, .\label{hmf2}
\end{eqnarray}
Again this decomposition is orthogonal, all four terms on the RHS of both decompositions are mutually orthogonal wrt \eqref{inner}.
Note that one can regroup terms in the decompositions (\ref{hmf1}, \ref{hmf2}) to write
\begin{eqnarray}
\omega&=&d\phi+\psi\qquad\mbox{where}\quad \delta\psi=0\,,\quad \rn\psi=0 \, ,\label{excc}\\
\omega&=&\delta\rho+\sigma\qquad\mbox{where}\quad d\sigma=0\,,\quad \rt\sigma=0 \, .\label{cexc}
\end{eqnarray}
In words these decompositions (\ref{excc}, \ref{cexc}) read: "Every form can be uniquely decomposed into an exact part and a co-closed part with vanishing normal component on the boundary" and "Every form can be uniquely decomposed into a co-exact part and a closed part with vanishing tangent component on the boundary"

\paragraph{Dirichlet and Neumann fields}(\cite{Schwarz:1995} Thm 2.6.1 and Cor 2.6.2, \cite{Duff:51})
When the boundary of a manifold $M$ is non-empty the space of harmonic fields is infinite dimensional. The subspaces of harmonic fields satisfying Dirichlet or Neumann boundary conditions respectively are however finite dimensional and isomorphic to the homology groups of $M$. To make this precise we first define Dirichlet fields $\hat\kappa$ and Neumann fields $\check \kappa$ as forms satisfying respectively
\begin{equation}
d\hat\kappa=\delta\hat\kappa=0\quad \mbox{with}\quad \rt\hat\kappa=0\qquad \mbox{and}\qquad
d\check\kappa=\delta\check\kappa=0 \quad \mbox{with}\quad \rn\check\kappa=0\,.
\end{equation}
The space of Dirichlet/Neumann fields of degree $p$ is then indicated with $\calh_{\mathrm{D}/\mathrm{N}}^p(M)$ respectively. The standard homology of cycles -- i.e. chains without boundary -- up to boundaries is referred to as the absolute homology of $M$ and we indicate the respective spaces of degree $p$ with $H_p(M)$. Additionally on a manifold with boundary one can define relative homology, where a relative cycle is defined as a chain whose boundary lies entirely in $\partial M$ and a chain is a relative boundary if it can be completed into a boundary by a chain lying entirely in $\partial M$. We indicate the space of relative homology classes -- relative chains modulo relative boundaries -- with $H_p(M,\partial M)$. The key result is then that
\begin{equation}
\calh_\mathrm{N}^p(M)\cong H_p(M)\cong H_{d-p}(M,\partial M)\, ,\qquad \calh_\mathrm{D}^p(M)\cong H_p(M,\partial M)\cong H_{d-p}(M)\, .\label{top rel}
\end{equation}
The isomorphism between absolute and relative cohomology goes under the name of Lefschetz duality, which reduces to Poincaré duality when the boundary is empty.

Note that the isomorphisms above together with the decompositions (\ref{excc}, \ref{cexc}) imply that on $M$ a closed $p$-form is exact iff $H_p(M)=\{0\}$ and that a co-closed $p$-form is co-exact iff $H_{d-p}(M)=\{0\}$.

\paragraph{Some first order boundary value problems}
Here we select a few results on some particular first order boundary value problems on $M$ that we will use in the solution of the second order boundary value problem of our interest in the next subsection.
\begin{itemize}
	\item[BV1] (\cite{Schwarz:1995} Thm 3.2.5) Given a closed boundary form $\rt \xi$ there exists a harmonic field $\kappa$, unique up to an arbitrary Dirichlet field, such that $\rt \kappa=\rt\xi$.
	\item[BV2] (\cite{Schwarz:1995} Cor 3.2.6) Given a closed form $\eta$, orthogonal to Neumann fields, then there exists a form $\omega$, unique up to an arbitrary Neumann field, such that $d\omega=\eta$, $\delta\omega=0$ and $\rn \omega=0$.
	\item[BV3] (\cite{Schwarz:1995} Thm 3.3.3) Consider a closed boundary form $\rt \xi$ such that $\int_{\pM} \rt \xi\wedge\star \rn \hat{\kappa}$=0 for all Dirichlet fields $\hat \kappa$. Then there exists a closed form $\omega$ such that $\rn \omega=0$ and $\rt \omega=\rt \xi$.
\end{itemize}

\subsection{Boundary value problem}
The second order boundary value problem of our interest is given in \eqref{const1vec}, (\ref{boundcond}, \ref{bcond2}). We will show a solution exists and characterize its form in this section. In the form language of this appendix the problem and its solution translates to
\begin{itemize}
	\item[BV] Given a boundary one-form $\rt\zeta$, a one-form $\chi$ exists such that $\delta d\chi=0$,  $\rn\chi=0$ and  $\rt\chi=\rt\zeta$, it is unique up to exact forms $d\alpha$ that vanish on the boundary. This holds under the assumption that $M$ is homologically trivial.
\end{itemize}

In the main text we restrict our analysis to $M$ being homeomorphic to a (closed) ball. As $M$ is then contractible to a point it is homologically trivial. As reviewed in the previous subsection this implies that -- except for degree 0 and $d$ -- there are no non-zero Dirichlet and Neumann fields. We start by applying the HMF decomposition \eqref{hmf1} to $\chi$, using the absence of Neumann fields
\begin{equation}
\chi=d\phi+\delta\beta\qquad \mbox{where} \quad \rn\delta\beta=0 \, .
\end{equation}
Using this decomposition the boundary value problem for $\chi$ translates as
\begin{equation}
\delta d\delta\beta=0\qquad \mbox{with} \quad \rn d\phi=0 \quad \mbox{and}\quad \rt d\phi+\rt\delta\beta=\rt\zeta \, .
\end{equation}
We can solve the second order problem for $\delta\beta$ as two subsequent first order problems. Let us call $\theta=d\delta\beta$, then $\theta$ satisfies
\begin{equation}
\delta \theta=d\theta=0 \quad \mbox{with} \quad \rt \theta=\rt d\zeta \, .
\end{equation}
By BV1 of the previous subsection, and using vanishing of all Dirichlet fields, it follows that $\theta$ exists and is unique. Given this unique $\theta$ we can then write
\begin{equation}
d\delta\beta=\theta,\quad \delta\delta\beta=0 \quad \mbox{with}\quad \rn\delta\beta=0 \, .
\end{equation}
This has the form of BV2 of the previous subsection, and using vanishing of all Neumann fields, it follows that $\delta\beta$ exists and is unique. The last step is to determine that there exists a $d\phi$ to complete $\chi$. This exact part satisfies the differential problem
\begin{equation}
dd\phi=0 \quad \mbox{with} \quad \rn d\phi=0,\quad \rt d\phi=\rt\zeta-\rt\delta\beta\,,
\end{equation}
which has a guaranteed solution by BV3 of the last subsection, because $d(\rt\zeta-\rt\delta\beta)=0$ and all Dirichlet fields are zero on $M$. Note that $d\phi$ is not unique since adding any $d\alpha$ such that $\left.d\alpha\right|_\pM=0$ would also give a solution. That such $d\alpha$ exist follows again from BV3.

Finally let us comment that the above existence proof can be replaced with an explicit construction of the solution using Green's forms \cite{DuffSpencer:51}.

  \bibliographystyle{JHEP}
      \bibliography{vac_lit}
\end{document}